# P-Governance Technology: Using Big Data for Political Party Management

Moniruzzaman Bhuiyan[1], Rafikul Haque[2] and Mahdi H. Miraz[3,4]
[1]Institute of Information Technology, University of Dhaka, Bangladesh
`mb@du.ac.bd`
[2]Samsung R&D Institute Bangladesh LTD, Bangladesh
`rafikulhaque@yahoo.com`
[3]Department of Computer Science & Software Engineering, University of Hail, KSA
`m.miraz@uoh.edu.sa`
[4]Department of Computing, Glyndŵr University, Wrexham, UK
`m.miraz@glyndwr.ac.uk`

*Abstract-* **Information and Communication Technology (ICT) has been playing a pivotal role since the last decade in developing countries that brings citizen services to the doorsteps and connecting people. With this aspiration ICT has introduced several technologies of citizen services towards all categories of people. The purpose of this study is to examine the Governance technology perspectives for political party, emphasizing on the basic critical steps through which it could be operationalized. We call it P-Governance. P-Governance shows technologies to ensure governance, management, interaction communication in a political party by improving decision making processes using big data. P-Governance challenges the competence perspective to apply itself more assiduously to operationalization, including the need to choose and give definition to one or more units of analysis (of which the routine is a promising candidate). This paper is to focus on research challenges posed by competence to which P-Governance can and should respond include different strategy issues faced by particular sections. Both the qualitative as well as quantitative research approaches were conducted. The standard of citizen services, choice & consultation, courtesy & consultation, entrance & information, and value for money have found the positive relation with citizen's satisfaction. This study results how can be technology make important roles on political movements in developing countries using big data.**

I. INTRODUCTION

An ancient Greek word "kebernon", that means to steer, has been followed for the term governance. Still in current world, governance means to the power that steering our political system, social system as well as all processes those have been used to control, to direct and to held an organization to their society. P-Governance is a new pattern and new processes to ensure the transparency, accountability, responsiveness, efficiency, supervision for all directions of any political party. Debating to decide policy and investment, to organize and to deliver information and services to the citizens P-Governance has opened new opportunities for citizens, supporters, party members, stakeholders etc.

Because of the increasing of world population from 1.65 billion to 6.79 billion within 20 years in 2010 [1], there are a lot of odd progresses have been increased also, such as primary need, those are causing poverty and increase of desperation among the nations. Governance for political parties comprises with effectiveness, equity and inclusiveness, participation and responsiveness, transparency and accountability, [1] among these population. So, while the population is larger, then there are more data around a political party to governance and empower them in proper management processes. Information and Communication Technology (ICT) can provide more opportunities towards the political parties to guide them in critical as well as general strategies and decision making which belongs to them. In Bangladesh, there are 85 million people out of 164.4 million involved with different political parties, where this population is more than British total population [18, 19]. There are more than 110 million mobile phone users, but ICT are not actively introduced in the process.

To represent and deliver social services with decentralization and to make policy framework [2] by proactive societal actors [3] in various public administrations, P-Governance technology is a better solution performing these functions and connecting citizens with political parties. Efficiency and effectiveness, stability of a political party, rule of law, citizen participations, corruption controlling and accountability [4], these six perspectives for P-Governance are key factors to improve governance processes using ICT. The Interrelationship between citizens and formal political parties influences on this critical process to use ICT key factors, such as social media data, mobile applications, web services etc. Hence, the collected big amount of data needs to be processed through ICT tools or services to achieve more efficient information for the governance processes of a political party.

Moreover, a citizen centric p-governance, supported by ICT, enjoys increasing trust of citizens and ensures accountability of organizational transactions [23]. It also provides enhanced collaboration among departments and stakeholders, thereby enabling fast decision making and consensus [24]. Now-a-days, Internet plays a vital role on the governance processes to take feedback from citizens, party member, supporters, stakeholders, etc. Thus it involves rethinking the entire redistribution of services across all agencies from a citizen perspective. So, through a holistic approach of a true citizen, the opportunity of interaction between political party and citizens could bring us to the doorstep of a new model of governance technology.

## II. WHAT IS P-GOVERNANCE

P-Governance (P-Gov) is not made on particular decisions of some specific people. That is a management system based on ICT. Rather, P-Gov is about to identify someone, whereas a input right belongs to him for different types of decision and to make people accountable on these rule of law [5]. The achievement on a performance goal, also comes from a governance principle, implanted on a corporate environment to manage it using ICT. So, P-Gov represents a framework of decision making and accountabilities to enlarge the domain of political activities using of ICT for big data [5]. Through this technology, as shown in Fig. 1, political parties can deliver public and social services to the nation with a responsive, efficient, transparent, accountable and effective organization using big data.

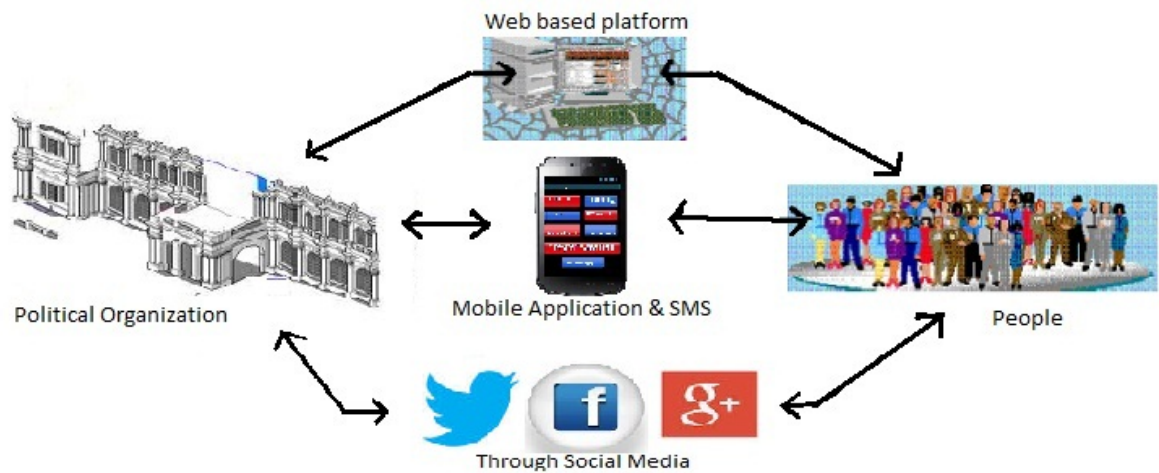

Figure 1. Concept of P-Governance for a political organization

The term "IT based governance" was introduced in 1970s [6] [7]. Then, in the late 1990s, e-government (e-Gov) was emerged. Because of the revolution of the Internet, the opportunity on business, governance, monitoring and tracking systems, effectiveness, participation, accountability etc has enlarged exponentially. The trend of a political personnel is to serve more efficient services to citizens, supporters, stakeholders etc as his customers by manipulating big data of each scope [8]. In this perspective, P-Governance has been started also for political personnel to fulfill the challenges as well as to practice their profession with root level people more promptly, smartly, deligatively and effectively for a nation.

III. Background

In the 2008 presidential election, the Democratic National Committee (DNC) of USA uses their mobilization programs among supporters, to cover up the participation among citizens, stakeholders, etc. and to pre-determine the statistics of election status. Obama uses Web based platforms, sharing through social media, and smart phones for his supporters to make them participate in the political processes of his election campaign for each voters [13].

P-Gov refers to governance technology for political parties, in the extension of e-Governance for political party management as well as the strategic use of political activities, which can be evaluate through web services, social media, or mobile services (using SMS or mobile app). It is about to use mobile applications and services, where an organization and political party can serve little more disciplined, transparent and accountable social infrastructure. Adapting to this infrastructure is almost not dispensable to fulfill the citizen's demand fully for all public sectors. It actually emerges as a huge opportunity in the using of ICT processes for public services, even if any additional thing added to the activities of political parties. These hopeful changes of technological and social sections are primary concern of our study. Through social media, a political organization can exchange their information with their citizens more interactively and vastly. So the way, through which a community or an individual can communicate, is now improved to a new horizon. Social media is more visible to our citizens directly through a popular platform on the Web. To register, mobilize, or persuade any supporter during campaign of a party, using a mobile application could be a better solution for them. Obama uses this technology for volunteer activity of most active supporters, canvassers, citizens, stakeholders to make them notice about his approach, to make them hear his speech and taking back a statistical report in return, such as rating, from them as suggestions without ringing the doorbell for their home during the campaigning period [13].

Critical Analysis regarding on P-Governance can be reported into four dimensions. Firstly, the strengths of this technology are Internet (as pull factor), combination with democratization reforms, willingness of that party, and modern image for that party. Secondly, the weakness of P-Governance could be lack of cyber laws in developing countries, hierarchy in organization, lack of expertise on technology, budget, lack of motivation, slow decision making process, integration to reform, short term approaches for election campaign. Thirdly, the opportunities, that could be our positive factors, are raising public sector fund through this technology, showing more competitive and transparency edge on natural change of processes towards the supporters, citizens and stakeholders. And fourthly, threats could be bureaucracy, piracy, misuse, corruption, maintain disorder, political disorder, resistance etc. for applying P-Governance technology.

IV. METHODOLOGY

The term big data refers to huge amount of datasets and analyzing big data refers to the steps of manipulating and testing these datasets to seek insight them. The idea of big data analyzing is a formal process of using open data around everywhere. It is not a new process in this world, but leading some habitudes related to amplify these datasets such as voting, rating, sharing, feeding activities of a political party through individual citizens. Big data sizes could be starting from a few terabytes and ending to a lot petabytes of datasets in a single field.

The six most critical and also focusable stages of operating a political organization through Information and Communication Technology (ICT) are as follows-

1. Participation and Responsiveness:

    Political participation in any activity means, facilitating greater citizen participation in policy decision-making processes using web technologies as well as to provide information and to support infrastructure for empowering citizens to gain their achievement. In this case, it is needed a qualitative responsive service among the citizens and the policy influencers and deplorers, stakeholders, supporters etc.

2. Transparency and Accountability:

    Through ICT a political party can be transparent in its infrastructure and activities towards the citizens, supporters, policy makers, stakeholders etc. This transparency makes more clear about the concept and ideal of that party to everyone. This increases the accountability and acceptance regarding any activity or performance of that political party to a society or a religion or a nation.

3. Effectiveness and efficiency:

    Exploring new channels of services, activities through ICT such as mobile service and social media for proactive two way communications ensures better effectiveness and efficiency for that political organization. Improving activity effectiveness for seeking better ways to interact with citizens and stakeholders using ICT, these political parties are aiming to become more efficient.

4. Equity and inclusiveness:

    By counseling and introducing ICT facilities such as web technologies, mobile services, social media in rural, remote and indigenous communities as well as creating more interest and inclusiveness on using ICT to these communities, may bring the equity to the citizens, supporters, party members, stakeholders etc. and will also help to focus the aim on political activities and their importance to everyone.

5. Rule of Law:

    Violation of the rule of law is not unusual in a democratic organization. The rule of law, that remains our guiding principles, violations for any political party such as corruption, illegal favors, etc. can be protected, monitored and/or preserved, and corrected using mobile services, social media, web technologies. Thus, using ICT any permanent structured or temporary unstructured groups or society can protest any violation of the rule of law of any political party over years.

6. Interrelation among citizens, political party, supporters, and the stakeholders:

    For adopting ICTs to modernize and for the reliability on political parties' a better interrelationship with citizens, supporters, and stakeholders offers considerable potential for the sustainable p-Governance. For

the improvements of political activities, policy decision-making, initiatives on broader perspectives by using of multiple channels (web technologies, social media, mobile services, etc.) to participate in local community life or online discussion forum, interrelationships play a vital role to for p-Governance.

The method of this technology is to extract relevant and interesting knowledge from big amount of data using data mining. Among several tasks of data mining classification [21] is one of the most useful techniques to build any model for decision making process from input data set. Then this model can be used to predict future data trends [22]. Hence, an application is developed for Android operating system as well as a web platform is also open for all citizens, supporters, stakeholders, party members etc. Two screen shots, as shown in Fig. 2, are presented here for the example of this development.

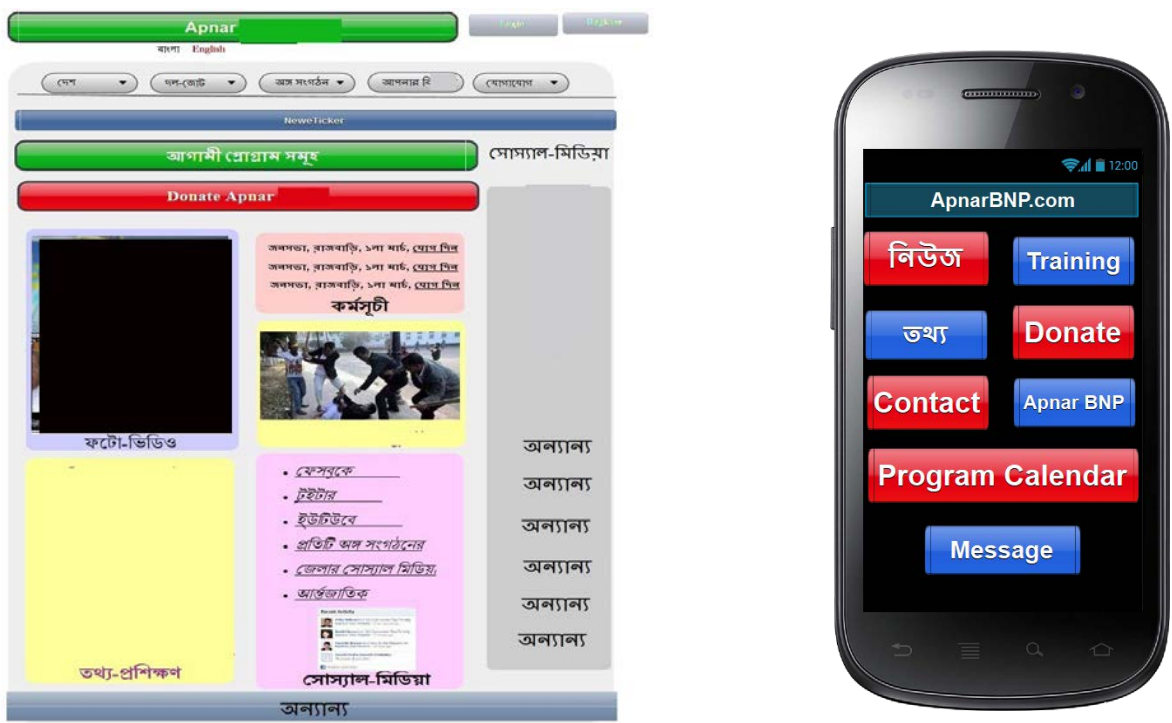

Figure 2. Web platform and mobile based application for a political party

The life cycle the processes executed to the following steps:

Step-1: A process could be started from a post in web platforms or in mobile based application. An authentic user can only in the application. A social media can also be included

Step-2: All users can view the post from web platforms or from mobile application. To suggest the post to another one, anyone can direct SMS the post through the mobile application or can suggest through social media.

Step-3: All users can comment on posts, share the post, or like the post as user review.

Step-4: A large amount of data can be collected for a single data set through prompt responses from huge users in the web based database.

Step-5: This data need to be classified for policy making processes, for identifying user experience, and for quick decision finalize. This process can be done through decision tree of "WEKA" tool from data sets.

Step-6: The already collected of data sets will be used as training data for identifying the future trend and possibility of user behaviors.

Step-7: A simplified result can be calculated through Bayesian belief networks, where attributes are conditionally independent not only to reduce the cost calculation, but also to count the class in many distributions.

$$P(C_j|V) \propto P(C_j)\prod_{i=1}^{n} P(v_i|C_j) \tag{1}$$

Relational model I: Naïve Bayes model classifier; where, $C_j$ is the j-th class, V is the data sample and $v_i$ is the value of i attributes from 1 to nth samples.

## V. SOLUTIONS

Regarding the critical analysis of P-Governance, there could be 4 phases to operate it between party to citizens (P2C), party to stakeholders (P2S), party to party (P2P). In the earliest phase, named information phase, all data will be collected accordingly in a common structure. In the next phase, named interaction phase, the communication with the citizen, stakeholders, party members will be operated through participating in discussion groups or forum, newsletter, etc. and notifying them using mail or service (mobile based or web based services) notifications or any other interactive processes. Hence, a huge collection of data is analyzed to take a quick decision on any activity. Data was collected from administered questions through many type of activities, those are performed, monitored, and guided by the supporters, members, stakeholders of different political organizations. The 4 phases, Information, Interaction, Transaction and Transformation could provide an acceptable solution to organize and manage the P-Governance technology for any political organization. The flow of processes of these phases, as shown in Fig. 3, is internally dependent with each other, which can be fully controlled through web based platforms, social media, mobile applications or services.

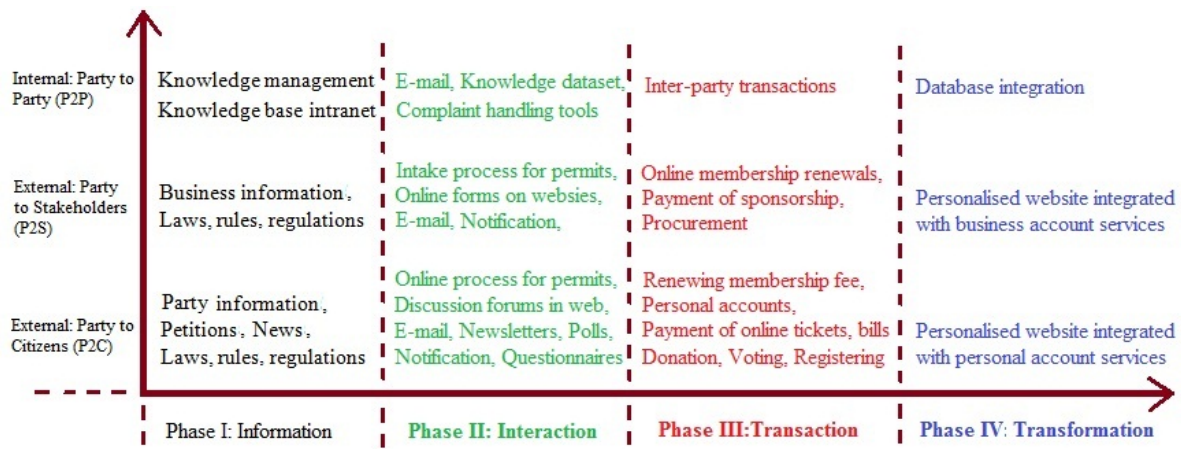

Figure 3. 4 Phases of the solution of P-Governance Technologies

## VI. Conceptual Framework

Reliability of P-Governance is widely understood as the use of the information and communication technology (ICT) to achieve:

- ➢ Providing information and delivering services,

- Participation and responsiveness for a citizen in the process of decision preparing,
- Transparent and accountable organization to members, citizens, supporters, stakeholders,
- More effective and efficient management.

The concept also seeks to improve the service delivery between Party-to-Citizens (P2C), Party-to-Stakeholders (P2S), Party-to-Party (P2P) as well as back office processes and interactions within the entire political framework to make it more effective and efficient management system. Parties all over the world are faced with the common challenge of improving their quality of services and gaining the confidence of their citizens. Citizen-centric service delivery has been identified as a key method to establish greater connections with the citizenry and build trust with them. This is to enable governments deliver better services to citizens more cost effectively.

The framework refers to 5 dimensions of P-Governance, as shown in Table-I, party, technology, society, supporters and interaction. For each of these dimensions, we may define a list of elements which are the part of the design of the framework for that dimension, e.g. target, value, operation, enforcement, role, service and institution within the dimension of party. These 5, as shown in Table 1, dimensions are the boundary spaces of the whole P-Governance framework, but the elements of these dimensions, which are the sole part of the design/framework, are open in all axes.

TABLE I
ELEMENTS OF 5 DIMENSIONS OF P- GOVERNANCE FRAMEWORK

| PARTY | TECHNOLOGY | SOCIETY | SUPPORTERS | INTERACTION |
|---|---|---|---|---|
| Target | Applications | Social inspection | Accessibility | Channel links |
| Value | Infrastructure | Public value | Delivery of services | Goals |
| Operation | Services | Globalization | Information sharing | Partnership |
| Enforcement | Data | Demography | Trust | Strategy |
| Role | Social Media | Transactions | Roles on changing | Governance |
| Service | Equipment | Digital inclusiveness | Responsiveness | |
| Institution | | | Participation | |

The framework also needs a common interface to communicate, participate, and responsiveness. The participation emphasizes on e-participation. In this case, governance could go through web technologies (already available), social media (partially implemented), and mobile apps services (rarely implemented). To input big data for a single data set and analyzing them for quick decision making as well as operating any activity, we need to merge them in one common platform.

VII. CONCLUSION

At the end of this study, this technology shows an immature research field which needs to be developed in multidisciplinary and a large practice. Though the definition of scope and core is still unclear and also a novelty, it brings a new analytical dimensions and new variables for governance technology research. This research needs more integration of ideas from political science, social science and psychological activities. So, P-Governance is a potentially fruitful research area. For any substantial change in this technology, it is important to recognize the research attention.

To make improve on this research, the state of this technology and its adjacent surroundings elements of different dimensions need to assess for building a good understanding to the citizens, supporters, stakeholders and members of any political party. To implement this research assessment framework, a political party need to focus on research methodology of 6 most critical stages, relational model for big data formulation, conceptual framework in 5 dimensions, solution in 4 phases for this technology, to collect and analyze data. This framework also needs to maintain disciplines to practice it for any political organization.

In conclusion, we can say that P-Governance is a key technology for any political party to minimize political disorders such as corruption, illegal favors etc. and to provide efficient and effective services to any society or nation.